\documentclass[aps, twocolumn, a4paper, superscriptaddress]{revtex4}

\usepackage{graphicx}
\usepackage{float}
\usepackage{color}
\usepackage{amsmath, amssymb}
\usepackage{lmodern}
\usepackage[T1]{fontenc}
\usepackage[utf8]{inputenc}
\usepackage[english]{babel}

\def\lsim{\mathrel{\rlap{\lower4pt\hbox{\hskip1pt$\sim$}}
    \raise1pt\hbox{$<$}}}       
\def\gsim{\mathrel{\rlap{\lower4pt\hbox{\hskip1pt$\sim$}}
    \raise1pt\hbox{$>$}}}       

\begin{document}

\title{Performance of weak species in the simplest generalization of the rock-paper-scissors model to four species}

\author{P.P. Avelino}
\affiliation{Instituto de Astrof\'{\i}sica e Ci\^encias do Espa{\c c}o, Universidade do Porto, CAUP, Rua das Estrelas, PT4150-762 Porto, Portugal}
\affiliation{Departamento de F\'{\i}sica e Astronomia, Faculdade de Ci\^encias, Universidade do Porto, Rua do Campo Alegre 687, PT4169-007 Porto, Portugal}
\affiliation{School of Physics and Astronomy, University of Birmingham, Birmingham B15 2TT, United Kingdom}
\author{B.F. de Oliveira}
\affiliation{Departamento de Física, Universidade Estadual de Maringá, Av. Colombo 5790, 87020-900 Maringá, PR, Brazil}
\author{R.S. Trintin}
\affiliation{Departamento de Física, Universidade Estadual de Maringá, Av. Colombo 5790, 87020-900 Maringá, PR, Brazil}

\begin{abstract}

We investigate the problem of the predominance and survival of "weak" species in the context of the simplest generalization of the spatial stochastic rock-paper-scissors model to four species by considering models in which one, two, or three species have a reduced predation probability. We show, using lattice based spatial stochastic simulations with random initial conditions, that if only one of the four species has its  probability reduced then the most abundant species is the prey of the "weakest" (assuming that the simulations are large enough for coexistence to prevail). Also, among the remaining cases, we present examples in which "weak" and "strong" species have similar average abundances and others in which either of them dominates --- the most abundant species being  always a prey of a "weak" species with which it maintains a unidirectional predator-prey interaction. However, in contrast to the three-species model, we find no systematic difference in the global performance of "weak" and "strong" species, and we conjecture that the same result will hold if the number of species is further increased. We also determine the probability of single species survival and coexistence as a function of the lattice size, discussing its dependence on initial conditions and on the change to the dynamics of the model which results from the extinction of one of the species.

\end{abstract}

\maketitle

\section{Introduction \label{sec1}}

Predator-prey models are a useful tool in the study of population dynamics in biological systems (see \cite{1920PNAS....6..410L,1926Natur.118..558V,May-Leonard} for the pioneer work by Lotka and Volterra, and May and Leonard). Among these, the spatial stochastic rock–paper–scissors (RPS) model describes the space-time evolution of three competing populations subject to cyclic non-hierarchical predator-prey interactions as well as reproduction and mobility. In the classical spatial stochastic RPS model \cite{2002-Kerr-N-418-171,Reichenbach-N-448-1046}, in which all the species have the same strength, the stable coexistence of all three species is generally possible if the mobility is not too large. Despite its simplicity, this model is able to successfully reproduce key dynamical features observed in simple biological systems with non-hierarchical selection  \cite{lizards,2002-Kerr-N-418-171,bacteria}.

The classical RPS model has been generalized to include additional species and interactions  \cite{2008-Peltomaki-PRE-78-031906,2008-Szabo-PRE-77-041919, 2011-Allesina-PNAS-108-5638, 2012-Avelino-PRE-86-031119, 2012-Avelino-PRE-86-036112,  2012-Li-PA-391-125, 2012-Roman-JSMTE-2012-p07014, 2013-Lutz-JTB-317-286, 2013-Roman-PRE-87-032148, 2014-Cheng-SR-4-7486, 2014-Szolnoki-JRSI-11-0735, 2016-Kang-Entropy-18-284, 2016-Roman-JTB-403-10, 2017-Brown-PRE-96-012147, 2017-Park-SR-7-7465, 2017-Bazeia-EPL-119-58003, 2017-Souza-Filho-PRE-95-062411, 2018-Shadisadt-PRE-98-062105, 2019-Avelino-PRE-99-052310}. Complex dynamical spatial structures (such as spirals with an arbitrary number of arms \cite{2012-Avelino-PRE-86-036112, 2017-Bazeia-EPL-119-58003, 2019-Bazeia-PRE-99-052408}, domain interfaces with or without  non-trivial internal dynamics \cite{2014-Avelino-PRE-89-042710}, and string networks with or without junctions \cite{2014-Avelino-PLA-378-393, 2017-Avelino-PLA-381-1014}), diverse scaling laws \cite{2012-Avelino-PRE-86-036112, 2017-Brown-PRE-96-012147}, and phase transitions \cite{2001-Szabo-PRE-63-061904, 2004-Szabo-PRE-69-031911, 2004-Szolnoki-PRE-70-037102, 2007-Perc-PRE-75-052102, 2007-Szabo-PRE-76-051921, 2008-Szabo-PRE-77-011906, 2011-Szolnoki-PRE-84-046106, 2013-Vukov-PRE-88-022123, 2018-Bazeia-EPL-124-68001} have been shown to arise naturally in many of these models. In most of them every species has the same strength, which results in the same average density for all species (if coexistence prevails) and a survival probability mainly dependent on initial conditions (in the absence of additional biases). 

In \cite{2001-Frean-PRSLB-268-1323} it has been shown that "weak" species have a competitive advantage in the context of a Lotka-Volterra implementation of the RPS model in which one of the three species --- usually refereed to as the "weakest" --- has a reduced predation probability. This problem has recently been revisited in the context of Lotka-Volterra and May-Leonard formulations of the spatial stochastic RPS model with random initial conditions \cite{2019PhRvE.100d2209A}. There, it has been shown that, despite the different population dynamics and spatial patterns, these two formulations lead to qualitatively similar results for the late time values of the relative abundances of the three species, as long as the simulation lattices are sufficiently large for coexistence to prevail — the "weakest" species generally having an advantage over the others (specially over its predator). On the other hand, in the case of small simulation lattices, a significant dependence of the probability of species survival on the lattice size has been found, associated to the relatively large oscillations taking place at the early stages of the simulations.

Here we study the problem of the predominance and survival of "weak" species in the simplest generalization of the spatial stochastic RPS model to an arbitrary number of species ($N_S$)  introduced in \cite{2012-Avelino-PRE-86-036112}. This model has been shown to give rise to a population network characterized by spiral patterns with $N_S$ arms, assuming that all the species have an equal strength. In this paper we relax this assumption, and investigate whether the positive impact of a reduced predation probability on species performance remains significant when the number of species is increased from three to four. 

\begin{figure}[H]
	\centering
	\includegraphics{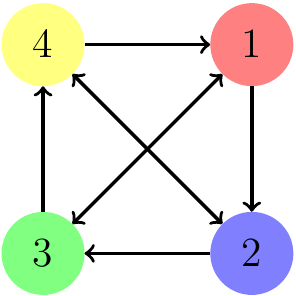}
	\caption{Scheme of the predator-prey interactions of our baseline RPS4 model.}
	\label{fig1}
\end{figure}

The  outline of this paper is as follows. We start by introducing the generalization of the spatial stochastic RPS model studied in the present paper as well as its numerical implementation in Sec. \ref{sec2}. In Sec. \ref{sec3} we present and discuss the results of a large number of spatial stochastic numerical simulations. Special emphasis is given to the way in which the average densities are affected by the reduced predation probabilities when coexistence prevails and to the dependence of the survival probability on the size of the simulation lattices. Finally, we conclude in Sec. \ref{conc}.

\section{Spatial Stochastic RPS4 model \label{sec2}}

In \cite{2012-Avelino-PRE-86-036112}, it has been shown, in the context of the simplest generalization of the spatial stochastic RPS model to $N_S$ species, that spirals with $N_S$ arms may arise in the context of competition models. Here, we shall focus on the May-Leonard formulation of the $4$-species sub-class of this family of models, which we shall refer to as RPS4. To this end, we shall consider a square lattice (see \cite{2004-Szabo-JPAMG-31-2599, 2009-Zhang-PRE-79-062901, 2014-Laird-Oikos-123-472, 2014-Rulquin-PRE-89-032133} for other lattice configurations) with $N^2$ sites and periodic boundary conditions  --- $N$ shall be referred to as its linear size. The different species are labelled by $i$ (or $j$) with $i,j = 1,...,4$, and modular arithmetic, where integers wrap around upon reaching $1$ or $4$, is assumed (the integers $i$ and $j$ represent the same species whenever $i = j \ {\rm mod} \ 4$, where mod denotes the modulo operation). 

In the May-Leonard formulation every site is either empty or occupied by a single individual of one of the four species. The number of individuals of the species $i$ and the number of empty sites will be denoted by $I_i$ and $I_0$, respectively --- the density of individuals of the species $i$ and the density of empty sites shall be defined by $\rho_i=I_i/N^2$ and $\rho_0 = I_0/N^2$, respectively. The possible interactions are predation 
\begin{equation}
i\ (i+1) \to i\ 0\,, \nonumber
\end{equation}
reproduction
\begin{equation}
 i\ 0 \to i\ i\,, \nonumber
\end{equation} 
and mobility 
\begin{equation}
 i\ \odot \to \odot\ i\,, \nonumber
\end{equation}
where $\odot$ represents either an individual of any species or an empty site. Reproduction and mobility interactions occur, respectively, with probabilities $r$ and $m$ (assumed to be the same for all the species). On the other hand, the predator-prey interactions of our baseline model are represented in Fig. \ref{fig1}, where the one-sided arrows represent one-directional predator-prey interactions between species $i$ and $i+1$, while the double sided arrows represent bi-directional predator-prey interactions between species $i$ and $i+2$. In our baseline model the predation probability $p$ is the same for all species. However, in this paper we shall investigate the dynamical impact of a reduction of the predation probability by a factor of $\mathcal{P}_w \in [0,1]$ of one, two or three of the four species.

At every simulation step, the algorithm randomly picks an occupied site to be the active one, randomly selects one of its adjacent neighbour sites to be the passive one, and randomly chooses an interaction to be executed by the individual at the active position: predation, mobility or reproduction with probabilities $p$, $m$ and $r$, respectively --- in this paper we use the von Neumann neighbourhood (or 4-neighbourhood) composed of a central cell (the active one) and its four non-diagonal adjacent cells (it has been shown in \cite{2019PhRvE.100d2209A}, in the context of a three species model, that a Moore neighbourhood leads to the same qualitative results). These three actions are repeated until a possible interaction is selected --- note that the interaction cannot be carried out whenever predation is selected and the passive is not a prey of the active, or if reproduction is selected and the passive is not an empty site. A generation time (our time unit) is defined as the time necessary for $N^2$ successive interactions to be completed. 

\begin{figure*}[t]
	\centering
    \includegraphics{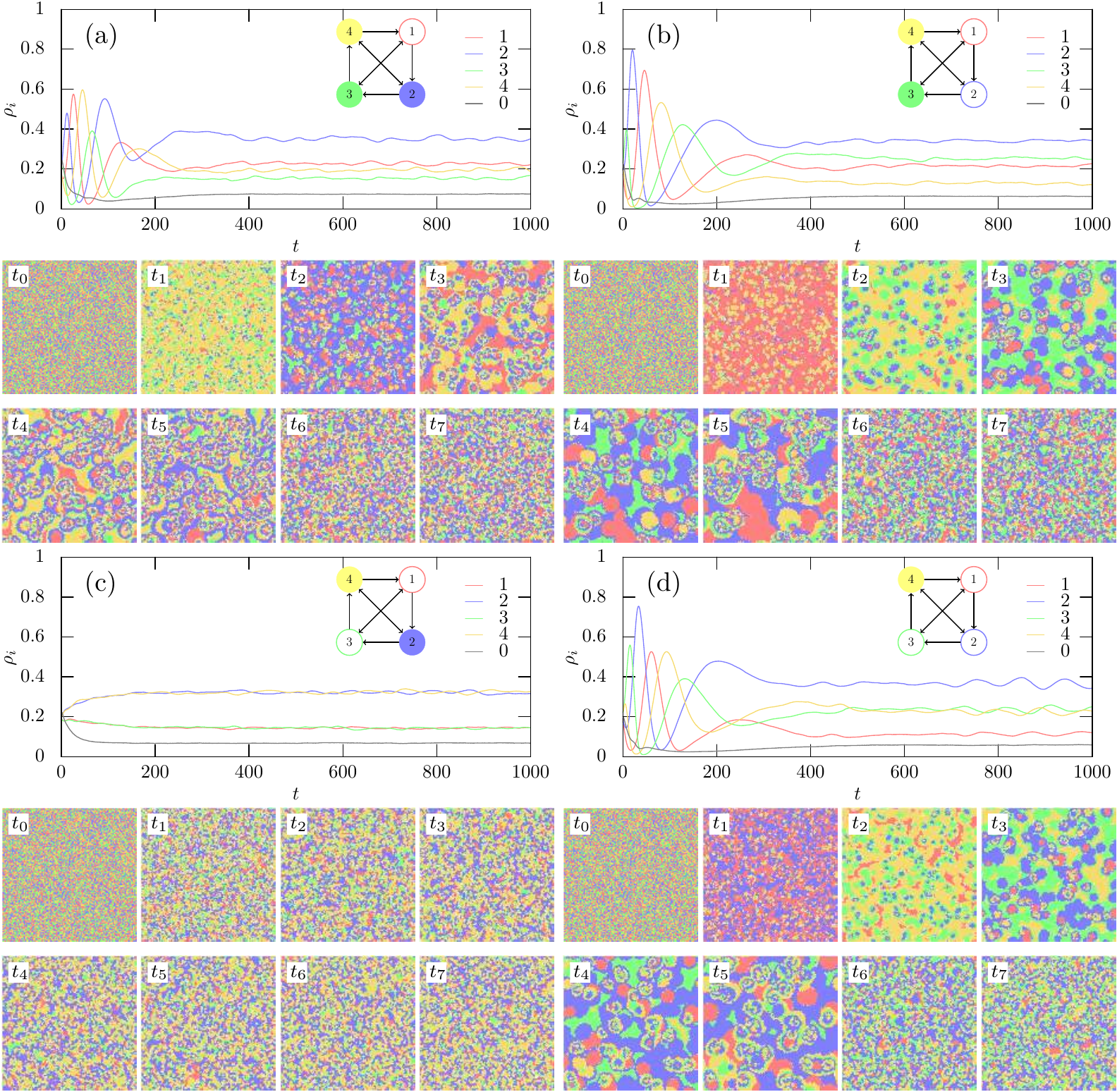}
	\caption{Graphs (a), (b), (c) and (d) show the evolution of the densities of the different species and empty sites ($\rho_i$ and $\rho_0$, respectively) over time for single realizations of the spatial stochastic RPS4 model (May-Leonard formulation), starting from random initial conditions with $\rho_1=\rho_2 = \rho_3 = \rho_4 = 1/4$. The model parameters are $m=0.2$, $p=r=0.4$, ${\mathcal P}_w=0.5$. The "strong" and "weak" species are represented, respectively, by a filled circle and a circumference. The lower panels of each graph show snapshots of the spatial distribution of the different species on a  $1000^2$ lattice at $t_0 = 0$, $t_1 = 50$, $t_2 = 100$, $t_3= 150$, $t_4 = 200$, $t_5 = 250$, $t_6 = 750$, and $t_7 = 5000$. Notice the changes in the background color at the early stages of simulations associated to rapid changes in the densities of the four species observed in graphs (a) (b) and (d).}
	\label{fig2}
\end{figure*}

\section{Results \label{sec3}}

In this section we shall describe the results of spatial stochastic numerical simulations  of the spatial RPS4 model in which one, two or three species have a reduced predation probability --- again, these species shall be referred to as "weak" and the others as "strong". 

The upper left (a), upper right (b), bottom left (c), and bottom right (d) graphs of Fig.  \ref{fig2} display the evolution of the densities of the different species and empty sites ($\rho_i$ and $\rho_0$, respectively) over time for single realizations of the spatial stochastic RPS4 model (May-Leonard formulation), starting from random initial conditions with $\rho_1=\rho_2 = \rho_3 = \rho_4 = 1/4$. The model parameters are $m=0.2$, $p=0.4$, $r=0.4$, ${\mathcal P}_w=0.5$ (a); $p_1=p{\mathcal P}_w$, $p_2=p_3=p_4=p$ (only species $1$ is "weak"); (b) $p_1=p_2=p{\mathcal P}_w$, $p_3=p_4=p$ (species $1$ and $2$ are "weak"); (c) $p_1=p_3=p{\mathcal P}_w$, $p_2=p_4=p$ (species $1$ and $3$ are "weak"); (d) $p_1=p_2=p_3=p{\mathcal P}_w$, $p_4=p$ (species $1$, $2$ and $3$ are "weak"), with the "strong" and "weak" species being represented, respectively, by a filled circle and a circumference. The lower panels of each graph show snapshots of the spatial distribution of the different species on a  $1000^2$ lattice at $t_0 = 0$, $t_1 = 50$, $t_2 = 100$, $t_3= 150$, $t_4 = 200$, $t_5 = 250$, $t_6 = 750$, and $t_7 = 5000$. Species $1$, $2$, $3$ and $4$ are represented in red, blue, green and yellow, respectively, while the empty sites are left in white. Notice the changes in the background color at the early stages of simulations associated to rapid changes in the densities of the four species observed in graphs (a) (b) and (d), before the steady-state configuration characterized by a distinctive spatial pattern consisting of a network of four-armed spirals is attained.

\begin{figure*}[t]
	\centering
	\includegraphics{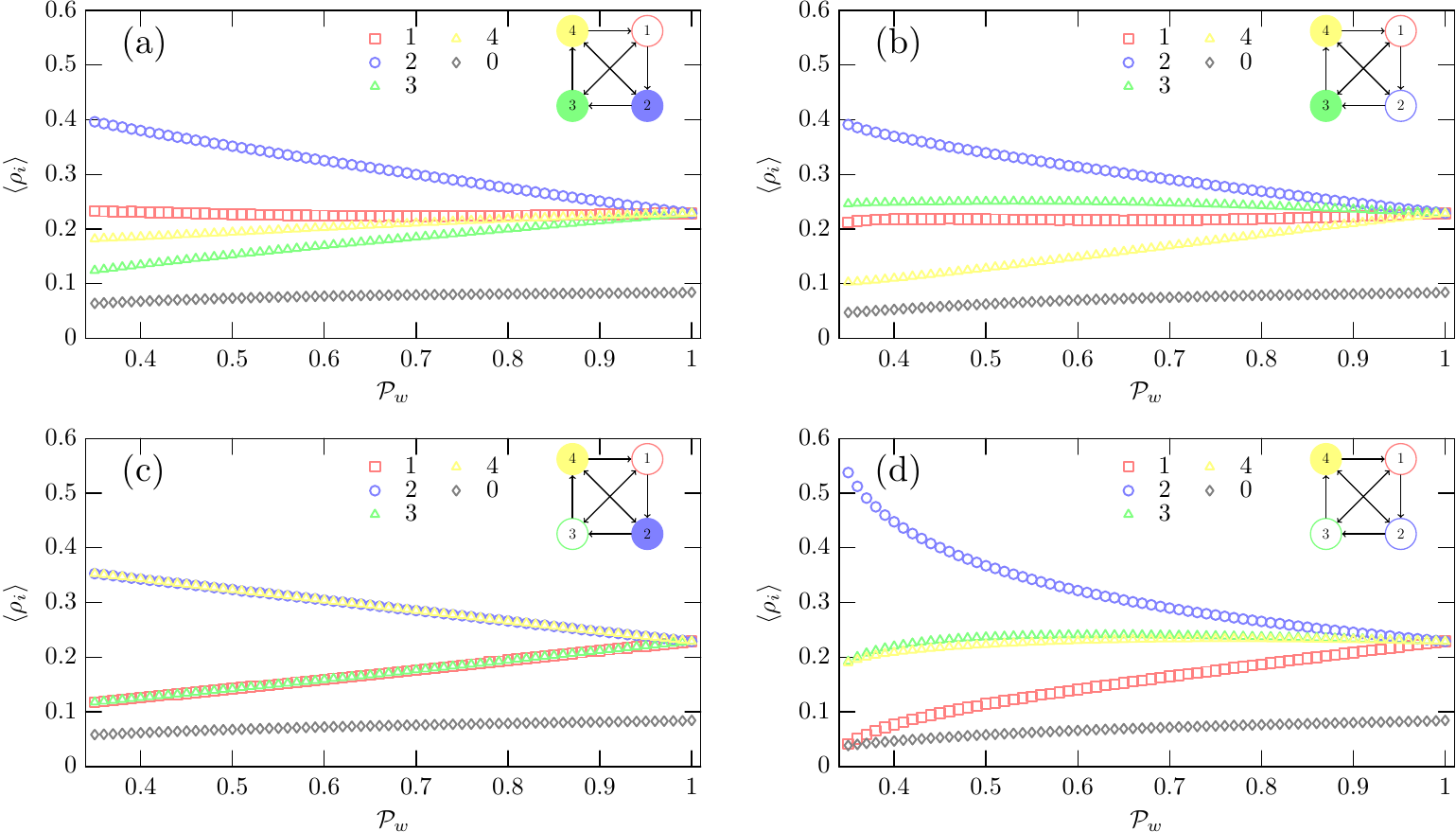}
	\caption{Average densities of the various species as a function of $\mathcal{P}_w$ assuming $m=0.2$, $p=r=0.4$. Each point results from an average over the last $10^4$ generations of $2000^2$ simulations with a time span equal to $1.5 \times 10^4$ generations. Notice that in cases (a) and (c) the most abundant species is "strong", while in cases (b) and (d) the most abundant species is "weak". Also, the most abundant species in all cases (the blue species) is always a prey of a "weak" species with which it maintains a unidirectional predator-prey interaction.}
	\label{fig3}
\end{figure*}

\begin{figure}[t]
	\centering
	\includegraphics{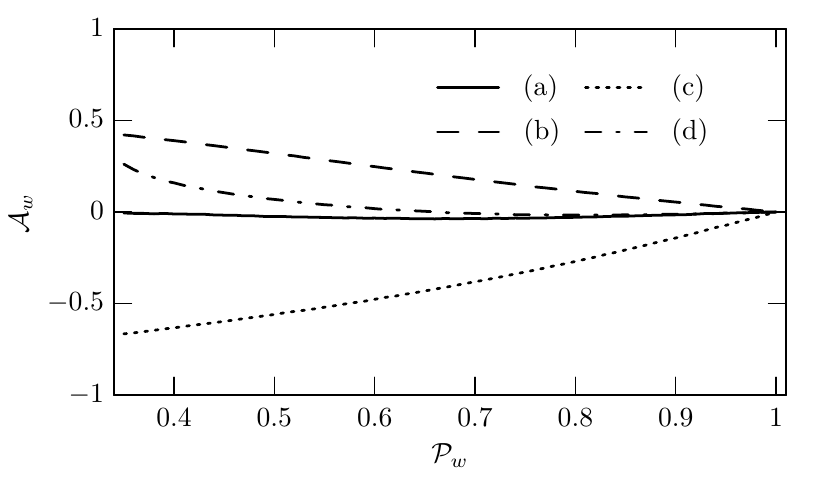}
	\caption{The relative advantage in being a weak species ${\mathcal A}_w$ (or disadvantage if $A_w < 0$) as a function of $\mathcal{P}_w$ for the cases (a), (b), (c) and (d) considered in Figs. \ref{fig2} and \ref{fig3}. Although the performance of "weak" and "strong "species" varies from case to case, their global average performance is not significantly different.}
	\label{fig4}
\end{figure}

Figure \ref{fig2} shows two cases, (b) and (d), in which one of the "weak" species is the most abundant and other two, (a) and (c), in which that does not happen. The two cases where there is a significant difference in the average abundance of "weak" and "species" are case (b), in which one of the "weak" species is the most abundant, and case (c), in which there is a significant advantage for both "strong" species. Nevertheless, Fig. \ref{fig2} already suggests that the average performance of "weak" and "strong" species is in general not very different if the simulations are sufficiently large for coexistence to prevail. Notice that in case (c), due to the model symmetry, the performance of the two "weak" species is identical (the same holding for the two "strong" species).

\begin{figure*}[t]
	\centering
	\includegraphics{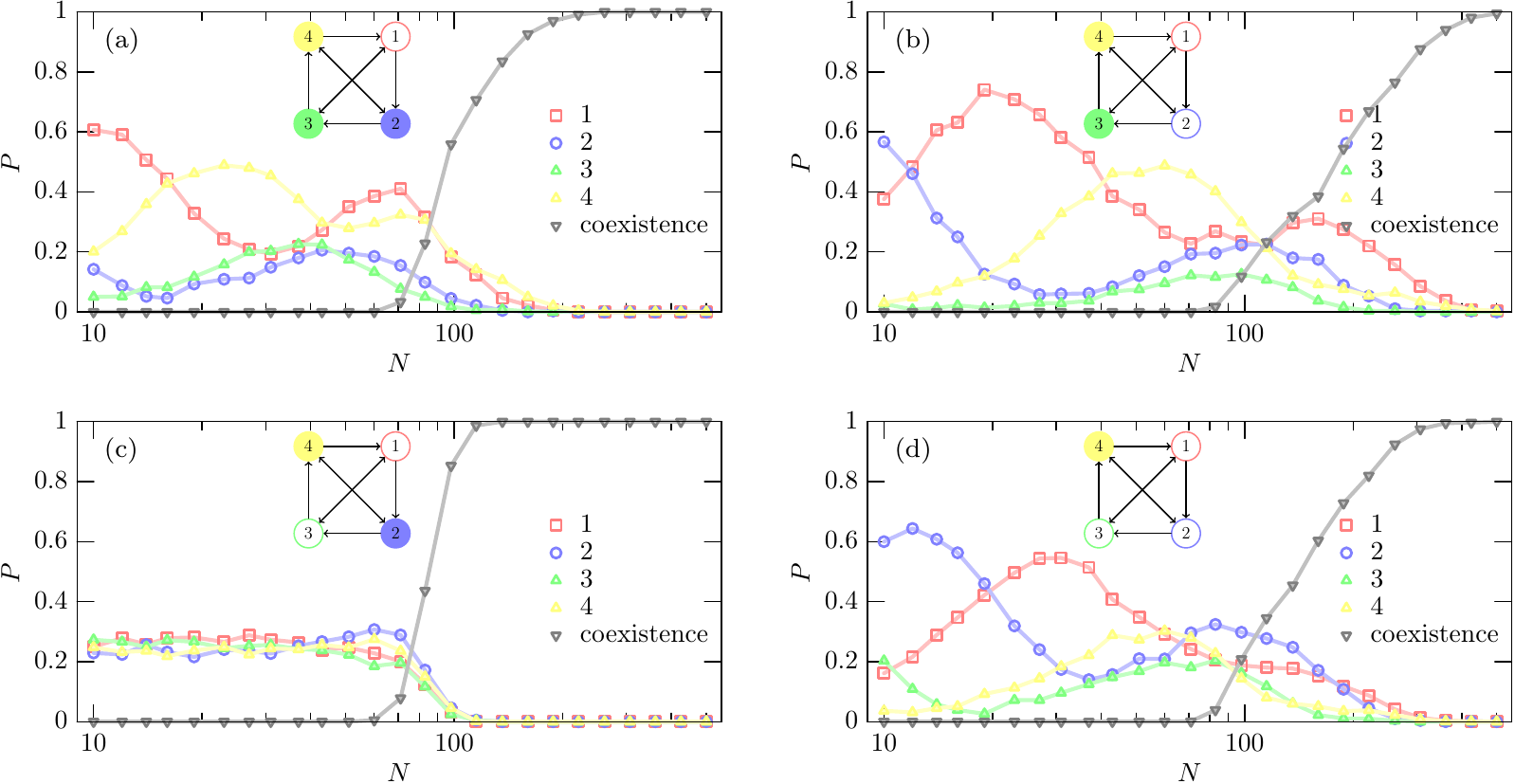}
	\caption{Probability $P$ of single species survival and coexistence as a function of the linear lattice size $N$ assuming the same parameters considered in Fig. \ref{fig2}. Each point was estimated from $10^3$ simulations with a total simulation time equal to $2 \times 10^4$ generations, starting from random initial conditions with $\rho_1=\rho_2 = \rho_3 = \rho_4 = 1/4$. The error bars are always smaller than the size of the symbols.}
	\label{fig5}
\end{figure*}

\begin{figure*}[t]
	\centering
	\includegraphics{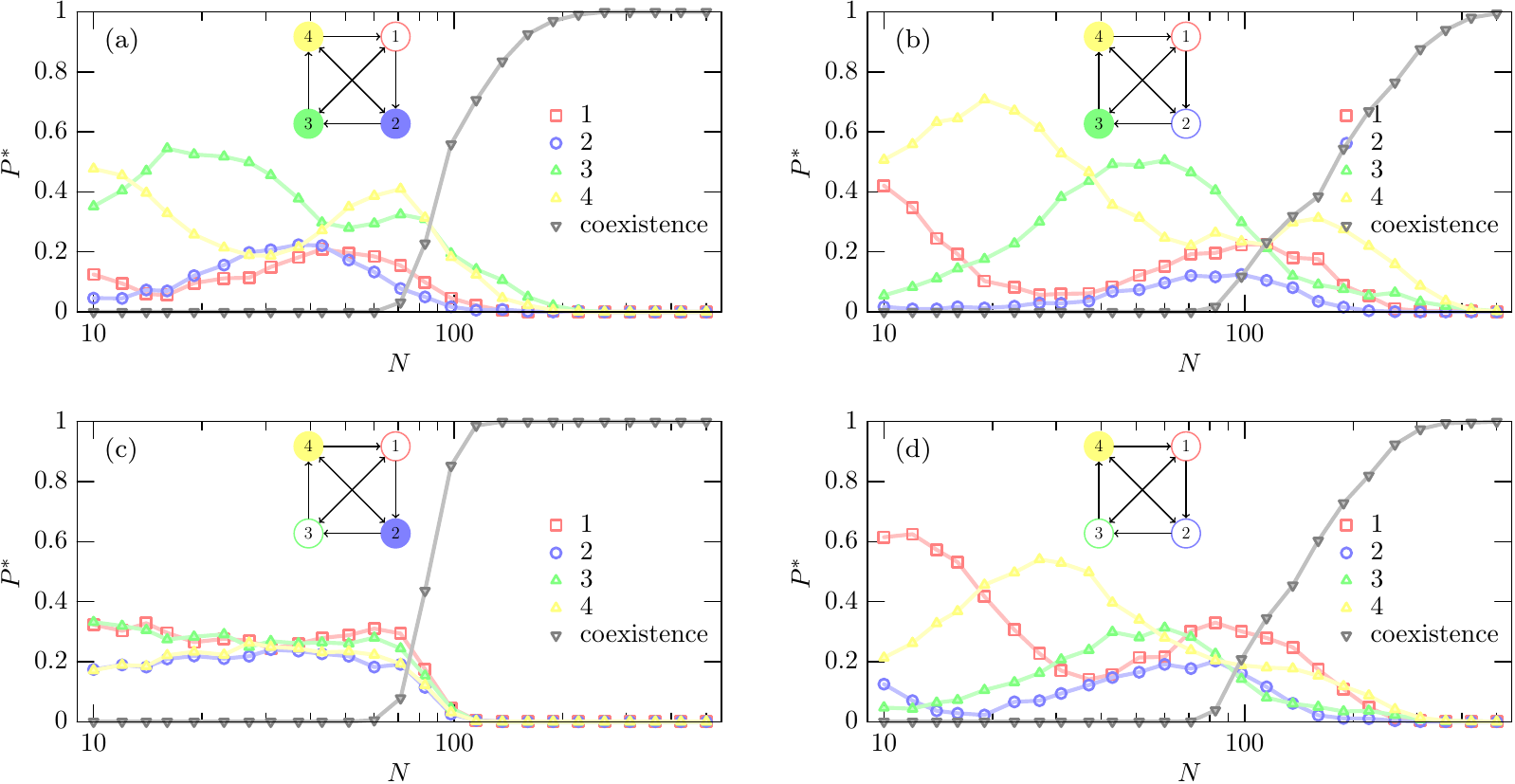}
	\caption{Probability $P^*$ that the species $i$ is the first species to become extinct, or that the coexistence of the four species is maintained, as a function of the linear lattice size $N$. $P^*$ was estimated using the same simulations considered in Fig. \ref{fig5}.}
	\label{fig6}
\end{figure*}

In order to investigate this aspect further we performed a large number of simulations of cases (a), (b), (c) and (d) with $m=0.2$, and $p=r=0.4$ (the same as in Fig. \ref{fig2}), but variable $\mathcal{P}_w$. Figure \ref{fig3} shows the value of the average density of the four species as a function of $\mathcal{P}_w$. The data points result from an average over the last $10^{4}$ generations of simulations with a time span equal to $1.5 \times 10^{4}$ generations performed on a $2000^{2}$ lattice, large enough to guarantee the preservation of coexistence in all simulations. The results for $\mathcal{P}_w=1$ were computed first, starting from random initial conditions. The final conditions of the simulations with $\mathcal{P}_w=1$ were then taken as initial conditions of new simulations with $\mathcal{P}_w=1-0.01$. The same  procedure was repeated until $\mathcal{P}_w=0.35$ was reached, thus ensuring a fast convergence of the simulations for every value of $\mathcal{P}_w$.

Figure \ref{fig3} shows that the most abundant species is "strong" in cases (a) and (c), and "weak" in cases (b) and (d). Again, note that, due to the model symmetry, and except for the different labelling, the two "strong" and the two "weak" species in case (c) are indistinguishable. In order to verify whether the species strength, on its own, is an advantage or disadvantage in terms of the overall abundance, we define the average density of "weak" and "strong" species as 
\begin{equation}
\langle \rho_w \rangle = \frac{1}{\#W} \sum_{i \in W} \langle \rho_i \rangle\,, \quad \langle \rho_s \rangle = \frac{1}{\#S} \sum_{i \in W} \langle \rho_i \rangle\,,
\end{equation}
where $W$ and $S$ are, respectively, the sets whose elements are the "weak" and "strong" species, and $\#$ is used to represent the number of elements of each set. Let us also define the parameter 
\begin{equation}
{\mathcal A}_w=\frac{\langle \rho_w \rangle-\langle \rho_s \rangle}{\max(|\langle \rho_w \rangle|,|\langle \rho_s \rangle|)}\,,
\end{equation}
whose absolute value represents the relative advantage (if ${\mathcal A}_w>0$) or disadvantage (if ${\mathcal A}_w<0$) in being a "weak" species. Figure \ref{fig4} shows the value of ${\mathcal A}_w$ as a function of $\mathcal{P}_w$ for the cases (a), (b), (c) and (d) considered in Fig. \ref{fig3}. It shows a case (case (a)) in which there is on average no advantage or disadvantage in being the "weakest" species, another (case (c)) in which the "weak" species have a significant disadvantage over the others, and other two in which the "weak" species have some advantage over the "strong" species ((d), specially for ${\mathcal P}_w \lsim 0.5$, and (b)). Globally these results show that the average performance of "weak" and "strong" species is not significantly different. Hence, we may conclude that the predominance of the "weak" species observed in RPS models with three species no longer holds when the number of species is increased to four, and we conjecture that the same will remain true if the number of species is further increased.

Figure \ref{fig5} displays the probability $P$ of single species survival and coexistence as a function of the linear lattice size $N$ for a May-Leonard formulation of the spatial stochastic RPS4 model with $m=0.2$, $p=0.4$, $r=0.4$, ${\mathcal P}_w=0.5$ and (a) $p_1=p{\mathcal P}_w$, $p_2=p_3=p_4=p$; (b) $p_1=p_2=p{\mathcal P}_w$, $p_3=p_4=p$; (c) $p_1=p_3=p{\mathcal P}_w$, $p_2=p_4=p$; (d) $p_1=p_2=p_3=p{\mathcal P}_w$, $p_4=p$. Each point was estimated from $10^3$ simulations with a total simulation time equal to $2 \times 10^4$ generations, starting from random initial conditions with $\rho_1=\rho_2 = \rho_3 = \rho_4 = 1/4$. The error bars are always smaller than the size of the symbols, with the one-sigma uncertainty in the value of $P$, at each point being approximately equal to $[P(1-P)/10^{3}]^{1/2}$, with a maximum of approximately $1.6 \times 10^{-2}$ for $P = 0.5$. 

Figure \ref{fig5} shows that the transient coherent oscillations of the abundances of the four species in the early stages of simulations (a), (b) and (d) observed in Fig. \ref{fig2}, are responsible for a significant dependence of the survival probability on the linear size of the lattices --- a feature also observed in the context of a three species RPS model in which one of the species has a reduced predation probability \cite{2019PhRvE.100d2209A}. Also, it is interesting to note that the species with the largest survival probability is not necessarily the most abundant species in Fig. \ref{fig3}, even if the probability of coexistence is high. In fact, in cases (b) and (d) the red species is only the third and fourth most abundant, respectively, as long as the linear size of the simulations is large enough for coexistence to prevail (see Fig. \ref{fig3}). However, for $N_{th}>100$ and $N_{th}>200$, respectively, the red species is also the one that with the highest survival probability in Fig. \ref{fig5}. The explanation of this apparent inconsistency resides on the fact that once one of the species disappears, there is a significant change in the nature of the model. 

In fact, above a given linear size threshold, the first species to become extinct is typically the least abundant in Fig. \ref{fig3}. Once that happens, the remaining species may then be classified as a function of a strength parameter $S_k$, with the subscript $k=-1,0$ or $1$ representing the number of preys minus the number of predators ($S_{-1} =i+3$, $S_0=i+2$, and $S_1=i+1$, where $i$ the species that is the first to become extinct). We verified that, in general, once species $i$ vanishes, species $S_{-1}=i+3$ and $S_0=i+2$ also become extinct (in that order), with the species $S_1=i+1$ (the prey of the first species to become extinct) being the one surviving in the end. 

This correspondence between the first species to become extinct and the surviving species may be confirmed by comparing Figs. \ref{fig5} and \ref{fig6} --- Fig. \ref{fig6} displays the probability $P^*$ that the species $i$ is the first species to become extinct, or that the coexistence of the four species is maintained, as a function of the linear lattice size $N$ for the same simulations considered in Fig. \ref{fig5}. Again, the error bars are always smaller than the size of the symbols, with the one-sigma uncertainty in the value of $P^*$, at each point being approximately equal to  $[P^{*}(1-P^{*})/10^{3}]^{1/2}$, with a maximum of approximately $1.6 \times 10^{-2}$ for $P^{*} = 0.5$. Under the transformation $i \to i+1$, Fig. \ref{fig6} would become very similar to Fig. \ref{fig5}, thus confirming our analysis. For example, in case (b), if the coexistence probability is high, one would expect that species $4$ (yellow: least abundant species in the case (b) shown in Fig. \ref{fig3}) would be the one with higher extinction probability (this may be confirmed in the top right panel of Fig. \ref{fig6} for $N>100$), thus implying that its prey (red species $1$) should be the one with the highest survival probability (this may be confirmed in the top right panel of Fig. \ref{fig5} for $N>100$).

We have also considered a modification of our model where there is no reduction to predation probabilities between species with bi-directional predator-prey interactions, and verified that this change has no significant impact on our results. 

\section{Conclusions \label{conc}}

In this paper we have added a new dimension to the problem of the predominance and survival of "weak" species, by investigating the simplest generalization of the spatial stochastic RPS model to four species in which one or more species have a reduced predation probability. We have shown, using lattice based spatial stochastic simulations of a May-Leonard model formulation, that if only one of the four species has a reduced predation probability it is the prey of the "weakest" that is the most abundant species, as long as the simulations are large enough for coexistence to be maintained. This is in contrast with the three species model where the "weakest" species is generally the most abundant. By considering cases with more than one "weak" species, we have also found that, unlike in the case of the three species model, there is no significant average advantage or disadvantage associated to being "weak" or "strong". We have also shown that in the RPS4 model, once one of the species becomes extinct, the surviving species is typically its prey, this result being largely independent of the number of "weak" and "strong" species and of the specific value of the parameter characterizing the reduction of the predation probability of the "weak" species.

\begin{acknowledgments}
P.P.A. acknowledges the support from Fundação para a Ciência e a Tecnologia (FCT) through the Sabbatical Grant No. SFRH/BSAB/150322/2019 and through the research grants UID/FIS/04434/2019, UIDB/04434/2020 and UIDP/04434/2020. B.F.O. and R.S.T. thank CAPES - Finance Code 001, Funda\c c\~ao Arauc\'aria, and INCT-FCx (CNPq/FAPESP) for financial and computational support.
\end{acknowledgments}


\begin{thebibliography}{45}
\expandafter\ifx\csname natexlab\endcsname\relax\def\natexlab#1{#1}\fi
\expandafter\ifx\csname bibnamefont\endcsname\relax
  \def\bibnamefont#1{#1}\fi
\expandafter\ifx\csname bibfnamefont\endcsname\relax
  \def\bibfnamefont#1{#1}\fi
\expandafter\ifx\csname citenamefont\endcsname\relax
  \def\citenamefont#1{#1}\fi
\expandafter\ifx\csname url\endcsname\relax
  \def\url#1{\texttt{#1}}\fi
\expandafter\ifx\csname urlprefix\endcsname\relax\def\urlprefix{URL }\fi
\providecommand{\bibinfo}[2]{#2}
\providecommand{\eprint}[2][]{\url{#2}}

\bibitem[{\citenamefont{{Lotka}}(1920)}]{1920PNAS....6..410L}
\bibinfo{author}{\bibfnamefont{A.~J.} \bibnamefont{{Lotka}}},
  \bibinfo{journal}{Proceedings of the National Academy of Science}
  \textbf{\bibinfo{volume}{6}}, \bibinfo{pages}{410} (\bibinfo{year}{1920}).

\bibitem[{\citenamefont{{Volterra}}(1926)}]{1926Natur.118..558V}
\bibinfo{author}{\bibfnamefont{V.}~\bibnamefont{{Volterra}}},
  \bibinfo{journal}{\nat} \textbf{\bibinfo{volume}{118}}, \bibinfo{pages}{558}
  (\bibinfo{year}{1926}).

\bibitem[{\citenamefont{May and Leonard}(1975)}]{May-Leonard}
\bibinfo{author}{\bibfnamefont{R.}~\bibnamefont{May}} \bibnamefont{and}
  \bibinfo{author}{\bibfnamefont{W.}~\bibnamefont{Leonard}},
  \bibinfo{journal}{SIAM Journal on Applied Mathematics}
  \textbf{\bibinfo{volume}{29}}, \bibinfo{pages}{243} (\bibinfo{year}{1975}).

\bibitem[{\citenamefont{Kerr et~al.}(2002)\citenamefont{Kerr, Riley, Feldman,
  and Bohannan}}]{2002-Kerr-N-418-171}
\bibinfo{author}{\bibfnamefont{B.}~\bibnamefont{Kerr}},
  \bibinfo{author}{\bibfnamefont{M.~A.} \bibnamefont{Riley}},
  \bibinfo{author}{\bibfnamefont{M.~W.} \bibnamefont{Feldman}},
  \bibnamefont{and} \bibinfo{author}{\bibfnamefont{B.~J.~M.}
  \bibnamefont{Bohannan}}, \bibinfo{journal}{Nature}
  \textbf{\bibinfo{volume}{418}}, \bibinfo{pages}{171} (\bibinfo{year}{2002}).

\bibitem[{\citenamefont{Reichenbach et~al.}(2007)\citenamefont{Reichenbach,
  Mobilia, and Frey}}]{Reichenbach-N-448-1046}
\bibinfo{author}{\bibfnamefont{T.}~\bibnamefont{Reichenbach}},
  \bibinfo{author}{\bibfnamefont{M.}~\bibnamefont{Mobilia}}, \bibnamefont{and}
  \bibinfo{author}{\bibfnamefont{E.}~\bibnamefont{Frey}},
  \bibinfo{journal}{Nature} \textbf{\bibinfo{volume}{448}},
  \bibinfo{pages}{1046} (\bibinfo{year}{2007}).

\bibitem[{\citenamefont{Sinervo and Lively}(1996)}]{lizards}
\bibinfo{author}{\bibfnamefont{B.}~\bibnamefont{Sinervo}} \bibnamefont{and}
  \bibinfo{author}{\bibfnamefont{C.~M.} \bibnamefont{Lively}},
  \bibinfo{journal}{Nature} \textbf{\bibinfo{volume}{380}},
  \bibinfo{pages}{240} (\bibinfo{year}{1996}).

\bibitem[{\citenamefont{Kirkup and Riley}(2004)}]{bacteria}
\bibinfo{author}{\bibfnamefont{B.~C.} \bibnamefont{Kirkup}} \bibnamefont{and}
  \bibinfo{author}{\bibfnamefont{M.~A.} \bibnamefont{Riley}},
  \bibinfo{journal}{Nature} \textbf{\bibinfo{volume}{428}},
  \bibinfo{pages}{412} (\bibinfo{year}{2004}).

\bibitem[{\citenamefont{Peltom\"{a}ki and
  Alava}(2008)}]{2008-Peltomaki-PRE-78-031906}
\bibinfo{author}{\bibfnamefont{M.}~\bibnamefont{Peltom\"{a}ki}}
  \bibnamefont{and} \bibinfo{author}{\bibfnamefont{M.}~\bibnamefont{Alava}},
  \bibinfo{journal}{Phys. Rev. E} \textbf{\bibinfo{volume}{78}},
  \bibinfo{pages}{031906} (\bibinfo{year}{2008}).

\bibitem[{\citenamefont{Szab{\'{o}} et~al.}(2008)\citenamefont{Szab{\'{o}},
  Szolnoki, and Borsos}}]{2008-Szabo-PRE-77-041919}
\bibinfo{author}{\bibfnamefont{G.}~\bibnamefont{Szab{\'{o}}}},
  \bibinfo{author}{\bibfnamefont{A.}~\bibnamefont{Szolnoki}}, \bibnamefont{and}
  \bibinfo{author}{\bibfnamefont{I.}~\bibnamefont{Borsos}},
  \bibinfo{journal}{Phys. Rev. E} \textbf{\bibinfo{volume}{77}},
  \bibinfo{pages}{041919} (\bibinfo{year}{2008}).

\bibitem[{\citenamefont{Allesina and
  Levine}(2011)}]{2011-Allesina-PNAS-108-5638}
\bibinfo{author}{\bibfnamefont{S.}~\bibnamefont{Allesina}} \bibnamefont{and}
  \bibinfo{author}{\bibfnamefont{J.~M.} \bibnamefont{Levine}},
  \bibinfo{journal}{PNAS} \textbf{\bibinfo{volume}{108}}, \bibinfo{pages}{5638}
  (\bibinfo{year}{2011}).

\bibitem[{\citenamefont{Avelino
  et~al.}(2012{\natexlab{a}})\citenamefont{Avelino, Bazeia, Losano, and
  Menezes}}]{2012-Avelino-PRE-86-031119}
\bibinfo{author}{\bibfnamefont{P.~P.} \bibnamefont{Avelino}},
  \bibinfo{author}{\bibfnamefont{D.}~\bibnamefont{Bazeia}},
  \bibinfo{author}{\bibfnamefont{L.}~\bibnamefont{Losano}}, \bibnamefont{and}
  \bibinfo{author}{\bibfnamefont{J.}~\bibnamefont{Menezes}},
  \bibinfo{journal}{Phys. Rev. E} \textbf{\bibinfo{volume}{86}},
  \bibinfo{pages}{031119} (\bibinfo{year}{2012}{\natexlab{a}}).

\bibitem[{\citenamefont{Avelino
  et~al.}(2012{\natexlab{b}})\citenamefont{Avelino, Bazeia, Losano, Menezes,
  and Oliveira}}]{2012-Avelino-PRE-86-036112}
\bibinfo{author}{\bibfnamefont{P.~P.} \bibnamefont{Avelino}},
  \bibinfo{author}{\bibfnamefont{D.}~\bibnamefont{Bazeia}},
  \bibinfo{author}{\bibfnamefont{L.}~\bibnamefont{Losano}},
  \bibinfo{author}{\bibfnamefont{J.}~\bibnamefont{Menezes}}, \bibnamefont{and}
  \bibinfo{author}{\bibfnamefont{B.~F.} \bibnamefont{Oliveira}},
  \bibinfo{journal}{Phys. Rev. E} \textbf{\bibinfo{volume}{86}},
  \bibinfo{pages}{036112} (\bibinfo{year}{2012}{\natexlab{b}}).

\bibitem[{\citenamefont{Li et~al.}(2012)\citenamefont{Li, Dong, and
  Yang}}]{2012-Li-PA-391-125}
\bibinfo{author}{\bibfnamefont{Y.}~\bibnamefont{Li}},
  \bibinfo{author}{\bibfnamefont{L.}~\bibnamefont{Dong}}, \bibnamefont{and}
  \bibinfo{author}{\bibfnamefont{G.}~\bibnamefont{Yang}},
  \bibinfo{journal}{Physica A: Statistical Mechanics and its Applications}
  \textbf{\bibinfo{volume}{391}}, \bibinfo{pages}{125} (\bibinfo{year}{2012}).

\bibitem[{\citenamefont{Roman et~al.}(2012)\citenamefont{Roman, Konrad, and
  Pleimling}}]{2012-Roman-JSMTE-2012-p07014}
\bibinfo{author}{\bibfnamefont{A.}~\bibnamefont{Roman}},
  \bibinfo{author}{\bibfnamefont{D.}~\bibnamefont{Konrad}}, \bibnamefont{and}
  \bibinfo{author}{\bibfnamefont{M.}~\bibnamefont{Pleimling}},
  \bibinfo{journal}{Journal of Statistical Mechanics: Theory and Experiment}
  \textbf{\bibinfo{volume}{2012}}, \bibinfo{pages}{P07014}
  (\bibinfo{year}{2012}).

\bibitem[{\citenamefont{L\"{u}tz et~al.}(2013)\citenamefont{L\"{u}tz,
  Risau-Gusman, and Arenzon}}]{2013-Lutz-JTB-317-286}
\bibinfo{author}{\bibfnamefont{A.~F.} \bibnamefont{L\"{u}tz}},
  \bibinfo{author}{\bibfnamefont{S.}~\bibnamefont{Risau-Gusman}},
  \bibnamefont{and} \bibinfo{author}{\bibfnamefont{J.~J.}
  \bibnamefont{Arenzon}}, \bibinfo{journal}{Journal of Theoretical Biology}
  \textbf{\bibinfo{volume}{317}}, \bibinfo{pages}{286} (\bibinfo{year}{2013}).

\bibitem[{\citenamefont{Roman et~al.}(2013)\citenamefont{Roman, Dasgupta, and
  Pleimling}}]{2013-Roman-PRE-87-032148}
\bibinfo{author}{\bibfnamefont{A.}~\bibnamefont{Roman}},
  \bibinfo{author}{\bibfnamefont{D.}~\bibnamefont{Dasgupta}}, \bibnamefont{and}
  \bibinfo{author}{\bibfnamefont{M.}~\bibnamefont{Pleimling}},
  \bibinfo{journal}{Phys. Rev. E} \textbf{\bibinfo{volume}{87}},
  \bibinfo{pages}{032148} (\bibinfo{year}{2013}).

\bibitem[{\citenamefont{Cheng et~al.}(2014)\citenamefont{Cheng, Yao, Huang,
  Park, Do, and Lai}}]{2014-Cheng-SR-4-7486}
\bibinfo{author}{\bibfnamefont{H.}~\bibnamefont{Cheng}},
  \bibinfo{author}{\bibfnamefont{N.}~\bibnamefont{Yao}},
  \bibinfo{author}{\bibfnamefont{Z.-G.} \bibnamefont{Huang}},
  \bibinfo{author}{\bibfnamefont{J.}~\bibnamefont{Park}},
  \bibinfo{author}{\bibfnamefont{Y.}~\bibnamefont{Do}}, \bibnamefont{and}
  \bibinfo{author}{\bibfnamefont{Y.-C.} \bibnamefont{Lai}},
  \bibinfo{journal}{Scientific Reports} \textbf{\bibinfo{volume}{4}},
  \bibinfo{pages}{7486} (\bibinfo{year}{2014}).

\bibitem[{\citenamefont{Szolnoki et~al.}(2014)\citenamefont{Szolnoki, Mobilia,
  Jiang, Szczesny, Rucklidge, and Perc}}]{2014-Szolnoki-JRSI-11-0735}
\bibinfo{author}{\bibfnamefont{A.}~\bibnamefont{Szolnoki}},
  \bibinfo{author}{\bibfnamefont{M.}~\bibnamefont{Mobilia}},
  \bibinfo{author}{\bibfnamefont{L.-L.} \bibnamefont{Jiang}},
  \bibinfo{author}{\bibfnamefont{B.}~\bibnamefont{Szczesny}},
  \bibinfo{author}{\bibfnamefont{A.~M.} \bibnamefont{Rucklidge}},
  \bibnamefont{and} \bibinfo{author}{\bibfnamefont{M.}~\bibnamefont{Perc}},
  \bibinfo{journal}{Journal of The Royal Society Interface}
  \textbf{\bibinfo{volume}{11}}, \bibinfo{pages}{20140735}
  (\bibinfo{year}{2014}).

\bibitem[{\citenamefont{Kang et~al.}(2016)\citenamefont{Kang, Pan, Wang, and
  He}}]{2016-Kang-Entropy-18-284}
\bibinfo{author}{\bibfnamefont{Y.}~\bibnamefont{Kang}},
  \bibinfo{author}{\bibfnamefont{Q.}~\bibnamefont{Pan}},
  \bibinfo{author}{\bibfnamefont{X.}~\bibnamefont{Wang}}, \bibnamefont{and}
  \bibinfo{author}{\bibfnamefont{M.}~\bibnamefont{He}},
  \bibinfo{journal}{Entropy} \textbf{\bibinfo{volume}{18}},
  \bibinfo{pages}{284} (\bibinfo{year}{2016}).

\bibitem[{\citenamefont{Roman et~al.}(2016)\citenamefont{Roman, Dasgupta, and
  Pleimling}}]{2016-Roman-JTB-403-10}
\bibinfo{author}{\bibfnamefont{A.}~\bibnamefont{Roman}},
  \bibinfo{author}{\bibfnamefont{D.}~\bibnamefont{Dasgupta}}, \bibnamefont{and}
  \bibinfo{author}{\bibfnamefont{M.}~\bibnamefont{Pleimling}},
  \bibinfo{journal}{Journal of Theoretical Biology}
  \textbf{\bibinfo{volume}{403}}, \bibinfo{pages}{10} (\bibinfo{year}{2016}).

\bibitem[{\citenamefont{Brown and Pleimling}(2017)}]{2017-Brown-PRE-96-012147}
\bibinfo{author}{\bibfnamefont{B.~L.} \bibnamefont{Brown}} \bibnamefont{and}
  \bibinfo{author}{\bibfnamefont{M.}~\bibnamefont{Pleimling}},
  \bibinfo{journal}{Phys. Rev. E} \textbf{\bibinfo{volume}{96}},
  \bibinfo{pages}{012147} (\bibinfo{year}{2017}).

\bibitem[{\citenamefont{Park et~al.}(2017)\citenamefont{Park, Do, Jang, and
  Lai}}]{2017-Park-SR-7-7465}
\bibinfo{author}{\bibfnamefont{J.}~\bibnamefont{Park}},
  \bibinfo{author}{\bibfnamefont{Y.}~\bibnamefont{Do}},
  \bibinfo{author}{\bibfnamefont{B.}~\bibnamefont{Jang}}, \bibnamefont{and}
  \bibinfo{author}{\bibfnamefont{Y.-C.} \bibnamefont{Lai}},
  \bibinfo{journal}{Scientific Reports} \textbf{\bibinfo{volume}{7}},
  \bibinfo{pages}{7465} (\bibinfo{year}{2017}).

\bibitem[{\citenamefont{Bazeia et~al.}(2017)\citenamefont{Bazeia, Menezes,
  de~Oliveira, and Ramos}}]{2017-Bazeia-EPL-119-58003}
\bibinfo{author}{\bibfnamefont{D.}~\bibnamefont{Bazeia}},
  \bibinfo{author}{\bibfnamefont{J.}~\bibnamefont{Menezes}},
  \bibinfo{author}{\bibfnamefont{B.~F.} \bibnamefont{de~Oliveira}},
  \bibnamefont{and} \bibinfo{author}{\bibfnamefont{J.~G. G.~S.}
  \bibnamefont{Ramos}}, \bibinfo{journal}{EPL} \textbf{\bibinfo{volume}{119}},
  \bibinfo{pages}{58003} (\bibinfo{year}{2017}).

\bibitem[{\citenamefont{Souza-Filho et~al.}(2017)\citenamefont{Souza-Filho,
  Bazeia, and Ramos}}]{2017-Souza-Filho-PRE-95-062411}
\bibinfo{author}{\bibfnamefont{C.~A.} \bibnamefont{Souza-Filho}},
  \bibinfo{author}{\bibfnamefont{D.}~\bibnamefont{Bazeia}}, \bibnamefont{and}
  \bibinfo{author}{\bibfnamefont{J.~G. G.~S.} \bibnamefont{Ramos}},
  \bibinfo{journal}{Phys. Rev. E} \textbf{\bibinfo{volume}{95}},
  \bibinfo{pages}{062411} (\bibinfo{year}{2017}).

\bibitem[{\citenamefont{Esmaeili et~al.}(2018)\citenamefont{Esmaeili, Brown,
  and Pleimling}}]{2018-Shadisadt-PRE-98-062105}
\bibinfo{author}{\bibfnamefont{S.}~\bibnamefont{Esmaeili}},
  \bibinfo{author}{\bibfnamefont{B.~L.} \bibnamefont{Brown}}, \bibnamefont{and}
  \bibinfo{author}{\bibfnamefont{M.}~\bibnamefont{Pleimling}},
  \bibinfo{journal}{Phys. Rev. E} \textbf{\bibinfo{volume}{98}},
  \bibinfo{pages}{062105} (\bibinfo{year}{2018}).

\bibitem[{\citenamefont{Avelino et~al.}(2019)\citenamefont{Avelino, Menezes,
  de~Oliveira, and Pereira}}]{2019-Avelino-PRE-99-052310}
\bibinfo{author}{\bibfnamefont{P.~P.} \bibnamefont{Avelino}},
  \bibinfo{author}{\bibfnamefont{J.}~\bibnamefont{Menezes}},
  \bibinfo{author}{\bibfnamefont{B.~F.} \bibnamefont{de~Oliveira}},
  \bibnamefont{and} \bibinfo{author}{\bibfnamefont{T.~A.}
  \bibnamefont{Pereira}}, \bibinfo{journal}{Phys. Rev. E}
  \textbf{\bibinfo{volume}{99}}, \bibinfo{pages}{052310}
  (\bibinfo{year}{2019}).

\bibitem[{\citenamefont{Bazeia et~al.}(2019)\citenamefont{Bazeia, de~Oliveira,
  and Szolnoki}}]{2019-Bazeia-PRE-99-052408}
\bibinfo{author}{\bibfnamefont{D.}~\bibnamefont{Bazeia}},
  \bibinfo{author}{\bibfnamefont{B.~F.} \bibnamefont{de~Oliveira}},
  \bibnamefont{and} \bibinfo{author}{\bibfnamefont{A.}~\bibnamefont{Szolnoki}},
  \bibinfo{journal}{Phys. Rev. E} \textbf{\bibinfo{volume}{99}},
  \bibinfo{pages}{052408} (\bibinfo{year}{2019}).

\bibitem[{\citenamefont{Avelino
  et~al.}(2014{\natexlab{a}})\citenamefont{Avelino, Bazeia, Losano, Menezes,
  and de~Oliveira}}]{2014-Avelino-PRE-89-042710}
\bibinfo{author}{\bibfnamefont{P.~P.} \bibnamefont{Avelino}},
  \bibinfo{author}{\bibfnamefont{D.}~\bibnamefont{Bazeia}},
  \bibinfo{author}{\bibfnamefont{L.}~\bibnamefont{Losano}},
  \bibinfo{author}{\bibfnamefont{J.}~\bibnamefont{Menezes}}, \bibnamefont{and}
  \bibinfo{author}{\bibfnamefont{B.~F.} \bibnamefont{de~Oliveira}},
  \bibinfo{journal}{Phys. Rev. E} \textbf{\bibinfo{volume}{89}},
  \bibinfo{pages}{042710} (\bibinfo{year}{2014}{\natexlab{a}}).

\bibitem[{\citenamefont{Avelino
  et~al.}(2014{\natexlab{b}})\citenamefont{Avelino, Bazeia, Menezes, and
  de~Oliveira}}]{2014-Avelino-PLA-378-393}
\bibinfo{author}{\bibfnamefont{P.~P.} \bibnamefont{Avelino}},
  \bibinfo{author}{\bibfnamefont{D.}~\bibnamefont{Bazeia}},
  \bibinfo{author}{\bibfnamefont{J.}~\bibnamefont{Menezes}}, \bibnamefont{and}
  \bibinfo{author}{\bibfnamefont{B.~F.} \bibnamefont{de~Oliveira}},
  \bibinfo{journal}{Physics Letters A} \textbf{\bibinfo{volume}{378}},
  \bibinfo{pages}{393} (\bibinfo{year}{2014}{\natexlab{b}}).

\bibitem[{\citenamefont{Avelino et~al.}(2017)\citenamefont{Avelino, Bazeia,
  Losano, Menezes, and de~Oliveira}}]{2017-Avelino-PLA-381-1014}
\bibinfo{author}{\bibfnamefont{P.~P.} \bibnamefont{Avelino}},
  \bibinfo{author}{\bibfnamefont{D.}~\bibnamefont{Bazeia}},
  \bibinfo{author}{\bibfnamefont{L.}~\bibnamefont{Losano}},
  \bibinfo{author}{\bibfnamefont{J.}~\bibnamefont{Menezes}}, \bibnamefont{and}
  \bibinfo{author}{\bibfnamefont{B.~F.} \bibnamefont{de~Oliveira}},
  \bibinfo{journal}{Physics Letters A} \textbf{\bibinfo{volume}{381}},
  \bibinfo{pages}{1014 } (\bibinfo{year}{2017}).

\bibitem[{\citenamefont{Szab\'o and
  Cz\'ar\'an}(2001)}]{2001-Szabo-PRE-63-061904}
\bibinfo{author}{\bibfnamefont{G.}~\bibnamefont{Szab\'o}} \bibnamefont{and}
  \bibinfo{author}{\bibfnamefont{T.}~\bibnamefont{Cz\'ar\'an}},
  \bibinfo{journal}{Phys. Rev. E} \textbf{\bibinfo{volume}{63}},
  \bibinfo{pages}{061904} (\bibinfo{year}{2001}).

\bibitem[{\citenamefont{Szab\'o and
  Arial~Sznaider}(2004)}]{2004-Szabo-PRE-69-031911}
\bibinfo{author}{\bibfnamefont{G.}~\bibnamefont{Szab\'o}} \bibnamefont{and}
  \bibinfo{author}{\bibfnamefont{G.}~\bibnamefont{Arial~Sznaider}},
  \bibinfo{journal}{Phys. Rev. E} \textbf{\bibinfo{volume}{69}},
  \bibinfo{pages}{031911} (\bibinfo{year}{2004}).

\bibitem[{\citenamefont{Szolnoki and
  Szab{\'{o}}}(2004)}]{2004-Szolnoki-PRE-70-037102}
\bibinfo{author}{\bibfnamefont{A.}~\bibnamefont{Szolnoki}} \bibnamefont{and}
  \bibinfo{author}{\bibfnamefont{G.}~\bibnamefont{Szab{\'{o}}}},
  \bibinfo{journal}{Phys. Rev. E} \textbf{\bibinfo{volume}{70}},
  \bibinfo{pages}{037102} (\bibinfo{year}{2004}).

\bibitem[{\citenamefont{Perc et~al.}(2007)\citenamefont{Perc, Szolnoki, and
  Szab\'o}}]{2007-Perc-PRE-75-052102}
\bibinfo{author}{\bibfnamefont{M.}~\bibnamefont{Perc}},
  \bibinfo{author}{\bibfnamefont{A.}~\bibnamefont{Szolnoki}}, \bibnamefont{and}
  \bibinfo{author}{\bibfnamefont{G.}~\bibnamefont{Szab\'o}},
  \bibinfo{journal}{Phys. Rev. E} \textbf{\bibinfo{volume}{75}},
  \bibinfo{pages}{052102} (\bibinfo{year}{2007}).

\bibitem[{\citenamefont{Szab\'o et~al.}(2007)\citenamefont{Szab\'o, Szolnoki,
  and Sznaider}}]{2007-Szabo-PRE-76-051921}
\bibinfo{author}{\bibfnamefont{G.}~\bibnamefont{Szab\'o}},
  \bibinfo{author}{\bibfnamefont{A.}~\bibnamefont{Szolnoki}}, \bibnamefont{and}
  \bibinfo{author}{\bibfnamefont{G.~A.} \bibnamefont{Sznaider}},
  \bibinfo{journal}{Phys. Rev. E} \textbf{\bibinfo{volume}{76}},
  \bibinfo{pages}{051921} (\bibinfo{year}{2007}).

\bibitem[{\citenamefont{Szab{\'{o}} and
  Szolnoki}(2008)}]{2008-Szabo-PRE-77-011906}
\bibinfo{author}{\bibfnamefont{G.}~\bibnamefont{Szab{\'{o}}}} \bibnamefont{and}
  \bibinfo{author}{\bibfnamefont{A.}~\bibnamefont{Szolnoki}},
  \bibinfo{journal}{Phys. Rev. E} \textbf{\bibinfo{volume}{77}},
  \bibinfo{pages}{011906} (\bibinfo{year}{2008}).

\bibitem[{\citenamefont{Szolnoki et~al.}(2011)\citenamefont{Szolnoki,
  Szab{\'{o}}, and Czak{\'{o}}}}]{2011-Szolnoki-PRE-84-046106}
\bibinfo{author}{\bibfnamefont{A.}~\bibnamefont{Szolnoki}},
  \bibinfo{author}{\bibfnamefont{G.}~\bibnamefont{Szab{\'{o}}}},
  \bibnamefont{and}
  \bibinfo{author}{\bibfnamefont{L.}~\bibnamefont{Czak{\'{o}}}},
  \bibinfo{journal}{Phys. Rev. E} \textbf{\bibinfo{volume}{84}},
  \bibinfo{pages}{046106} (\bibinfo{year}{2011}).

\bibitem[{\citenamefont{Vukov et~al.}(2013)\citenamefont{Vukov, Szolnoki, and
  Szab{\'{o}}}}]{2013-Vukov-PRE-88-022123}
\bibinfo{author}{\bibfnamefont{J.}~\bibnamefont{Vukov}},
  \bibinfo{author}{\bibfnamefont{A.}~\bibnamefont{Szolnoki}}, \bibnamefont{and}
  \bibinfo{author}{\bibfnamefont{G.}~\bibnamefont{Szab{\'{o}}}},
  \bibinfo{journal}{Phys. Rev. E} \textbf{\bibinfo{volume}{88}},
  \bibinfo{pages}{022123} (\bibinfo{year}{2013}).

\bibitem[{\citenamefont{Bazeia et~al.}(2018)\citenamefont{Bazeia, de~Oliveira,
  and Szolnoki}}]{2018-Bazeia-EPL-124-68001}
\bibinfo{author}{\bibfnamefont{D.}~\bibnamefont{Bazeia}},
  \bibinfo{author}{\bibfnamefont{B.~F.} \bibnamefont{de~Oliveira}},
  \bibnamefont{and} \bibinfo{author}{\bibfnamefont{A.}~\bibnamefont{Szolnoki}},
  \bibinfo{journal}{{EPL} (Europhysics Letters)}
  \textbf{\bibinfo{volume}{124}}, \bibinfo{pages}{68001}
  (\bibinfo{year}{2018}).

\bibitem[{\citenamefont{Frean and Abraham}(2001)}]{2001-Frean-PRSLB-268-1323}
\bibinfo{author}{\bibfnamefont{M.}~\bibnamefont{Frean}} \bibnamefont{and}
  \bibinfo{author}{\bibfnamefont{E.~R.} \bibnamefont{Abraham}},
  \bibinfo{journal}{Proc. R. Soc. Lond. B} \textbf{\bibinfo{volume}{268}},
  \bibinfo{pages}{1323} (\bibinfo{year}{2001}).

\bibitem[{\citenamefont{{Avelino} et~al.}(2019)\citenamefont{{Avelino}, {de
  Oliveira}, and {Trintin}}}]{2019PhRvE.100d2209A}
\bibinfo{author}{\bibfnamefont{P.~P.} \bibnamefont{{Avelino}}},
  \bibinfo{author}{\bibfnamefont{B.~F.} \bibnamefont{{de Oliveira}}},
  \bibnamefont{and} \bibinfo{author}{\bibfnamefont{R.~S.}
  \bibnamefont{{Trintin}}}, \bibinfo{journal}{\pre}
  \textbf{\bibinfo{volume}{100}}, \bibinfo{eid}{042209} (\bibinfo{year}{2019}).

\bibitem[{\citenamefont{Szab{\'{o}} et~al.}(2004)\citenamefont{Szab{\'{o}},
  Szolnoki, and Izs{\'{a}}k}}]{2004-Szabo-JPAMG-31-2599}
\bibinfo{author}{\bibfnamefont{G.}~\bibnamefont{Szab{\'{o}}}},
  \bibinfo{author}{\bibfnamefont{A.}~\bibnamefont{Szolnoki}}, \bibnamefont{and}
  \bibinfo{author}{\bibfnamefont{R.}~\bibnamefont{Izs{\'{a}}k}},
  \bibinfo{journal}{Journal of Physics A: Mathematical and General}
  \textbf{\bibinfo{volume}{37}}, \bibinfo{pages}{2599} (\bibinfo{year}{2004}).

\bibitem[{\citenamefont{Zhang et~al.}(2009)\citenamefont{Zhang, Chen, Qi, and
  Qing}}]{2009-Zhang-PRE-79-062901}
\bibinfo{author}{\bibfnamefont{G.-Y.} \bibnamefont{Zhang}},
  \bibinfo{author}{\bibfnamefont{Y.}~\bibnamefont{Chen}},
  \bibinfo{author}{\bibfnamefont{W.-K.} \bibnamefont{Qi}}, \bibnamefont{and}
  \bibinfo{author}{\bibfnamefont{S.-M.} \bibnamefont{Qing}},
  \bibinfo{journal}{Phys. Rev. E} \textbf{\bibinfo{volume}{79}},
  \bibinfo{pages}{062901} (\bibinfo{year}{2009}).

\bibitem[{\citenamefont{Laird}(2014)}]{2014-Laird-Oikos-123-472}
\bibinfo{author}{\bibfnamefont{R.~A.} \bibnamefont{Laird}},
  \bibinfo{journal}{Oikos} \textbf{\bibinfo{volume}{123}}, \bibinfo{pages}{472}
  (\bibinfo{year}{2014}).

\bibitem[{\citenamefont{Rulquin and
  Arenzon}(2014)}]{2014-Rulquin-PRE-89-032133}
\bibinfo{author}{\bibfnamefont{C.}~\bibnamefont{Rulquin}} \bibnamefont{and}
  \bibinfo{author}{\bibfnamefont{J.~J.} \bibnamefont{Arenzon}},
  \bibinfo{journal}{Phys. Rev. E} \textbf{\bibinfo{volume}{89}},
  \bibinfo{pages}{032133} (\bibinfo{year}{2014}).

\end{thebibliography}
\end{document}